# The Cornell Caltech Atacama Telescope

Simon J. E. Radford[a], Riccardo Giovanelli[b], Thomas A. Sebring[b], Jonas Zmuidzinas[a]

[a] California Institute of Technology, 320-47, Pasadena, CA  91125
[b] Center for Radiophysics and Space Research, Cornell University, Ithaca, NY  14853

## ABSTRACT

Cornell University, the California Institute for Technology, and the Jet Propulsion Laboratory are jointly studying the construction of a 25 m diameter telescope for submillimeter astronomy on a high mountain in northern Chile. This Cornell Caltech Atacama Telescope (CCAT) will combine high sensitivity, a wide field of view, and a broad wavelength range to provide an unprecedented capability for deep, large area, multi-color submillimeter surveys to complement narrow field, high resolution studies with ALMA. CCAT observations will address fundamental themes in contemporary astronomy, notably the formation and evolution of galaxies, the nature of the dark matter and dark energy that comprise most of the content of the universe, the formation of stars and planets, the conditions in circumstellar disks, and the conditions during the early history of the Solar system. The candidate CCAT site, at 5600 m in northern Chile, enjoys superb observing conditions. To accommodate large format bolometer cameras, CCAT is designed with a 20 arcmin field of view. CCAT will incorporate closed loop active control of its segmented primary mirror to maintain a half wavefront error of 10 µm rms or less. Instrumentation under consideration includes both short (650 µm–200 µm) and long (2 mm–750 µm) wavelength bolometer cameras, direct detection spectrometers, and heterodyne receiver arrays. The University of Colorado, a Canadian university consortium, and the UK Astronomy Technology Centre on behalf of the UK community are pursuing participation in the CCAT consortium. When complete early in the next decade, CCAT will be the largest and most sensitive facility of its class as well as the highest altitude astronomical facility on Earth.

## 1.  INTRODUCTION

Led by pioneering research at the CSO, the JCMT, and other telescopes, submillimeter astronomy has made tremendous advances in the last twenty years. Two notable examples are the discovery of a population of optically inconspicuous but submillimeter luminous galaxies in the early universe with the SCUBA camera (Holland et al. 1999) and the recognition the integrated intensity of the far IR and submillimeter radiation in the universe equals the intensity at optical wavelengths (Hauser & Dwek 2001). At the same time, the size of background limited bolometer arrays for submillimeter observations is increasing rapidly, doubling every couple of years. These cameras, which are well suited to high sensitivity, wide field surveys, will soon outstrip the capabilities of existing telescopes. Bolometer arrays are complementary to the heterodyne technology used in interferometers, such as ALMA. Finally, meteorological and radiometric studies in the high Andes of northern Chile have identified superb sites there, better than Mauna Kea, for ground based submillimeter astronomy.

These factors have motivated Cornell University, the California Institute for Technology (Caltech), and the Jet Propulsion Laboratory (JPL) to jointly study the construction of a 25 m telescope for submillimeter astronomy on a high mountain in northern Chile. With a 20 arcmin field of view, this Cornell Caltech Atacama Telescope (CCAT; Figure 1) will emphasize sensitive wide field observations with large format bolometer cameras. CCAT will have a larger aperture, better quality optics, a larger field of view, and a better site than existing telescopes. By identifying an abundance of sources for later detailed study, CCAT's large area surveys will complement the narrow field, high spectral and spatial resolution capabilities of ALMA. The CCAT feasibility and concept design study (Sebring et al. 2006a) was completed at the end of 2005 and the project received a strong endorsement from an independent review committee in 2006 January. The CCAT consortium is now expanding and the project plan aims for initial observations in 2012.

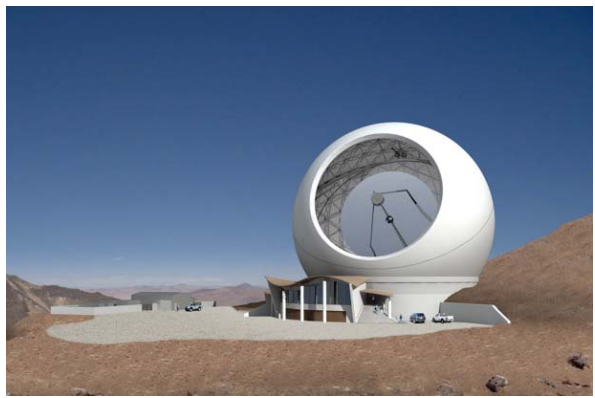

Fig. 1. CCAT at 5600 m on Cerro Chajnantor, Chile (concept view).

## 2. SCIENCE HIGHLIGHTS

Although hydrogen and helium make up over 98% of the baryonic matter in the Universe, in many cases it is the heavier elements, notably carbon, oxygen, silicon, and iron that allow us to discover and study distant objects. These elements form complex molecules and small dust particles that in many astrophysical environments obscure optical and ultraviolet photons and radiate predominantly at submillimeter wavelengths. Many of the most powerful and interesting phenomena in the universe, from star forming regions in our own galaxy to entire starburst galaxies in the early universe, are shrouded by dust and are completely inaccessible with optical observations. This makes submillimeter observations a particularly valuable probe of many astrophysical sources.

As well as being substantially larger and more sensitive than existing submillimeter telescopes, CCAT will be the first large submillimeter telescope designed specifically for wide field imaging. Hence it will provide an unparalleled ability to address key astronomical questions by mapping large areas of the sky. Science highlights include:

*Galaxy formation and evolution:* CCAT will detect hundreds of thousands of submillimeter starburst galaxies from the era of galaxy formation and assembly 10–12 billion years ago ($z = 2$–4) and will probe the earliest bursts of dusty star formation as far back as $z \sim 10$, less than 500 million years after the Big Bang. These observations will investigate the star formation history of the early universe, including the evolution of the population, the luminosity distribution, and the clustering of these galaxies.

*Dark Matter and dark energy:* CCAT's high resolution images of the Sunyaev-Zel'dovich effect in hundreds of clusters of galaxies will illustrate in detail how clusters form and evolve and will aid the determination of the dark energy equation of state and other cosmological parameters from SZ survey catalogs.

*Star Formation:* CCAT will provide the first complete census of cold, dense Galactic molecular cores that collapse to form stars. In nearby clouds, CCAT will detect 0.08 $M_\odot$ cores, smaller than the lowest mass stars.

*Protoplanetary and debris disks:* CCAT will survey nearby young star clusters to determine the prevalence and evolution of protoplanetary and debris disks. In conjunction with high resolution ALMA imaging, CCAT observations will study disk structure.

*The Kuiper belt:* Beyond Neptune, the Kuiper belt is a relic containing a record of the processes active in the early solar system, i. e., the accretion, migration, and clearing phases. CCAT will determine basic physical data – sizes and albedos – for hundreds of Kuiper belt objects, helping to anchor models of planet formation in the early solar system.

## 3. TELESCOPE SITE

Over the past decade, several groups have evaluated conditions for submillimeter astronomy at sites at and above 5000 m in the Atacama region of northern Chile (Radford & Holdaway 1998; Giovanelli et al. 2001). The measurements demonstrate these sites enjoy excellent observing conditions with extremely low atmospheric water vapor content. Observing conditions are considerably better than Mauna Kea and are comparable to the Antarctic plateau. As a result several telescopes have already been established in the vicinity of the village of San Pedro de Atacama, notably the international ALMA project now under construction on the 5000 m Chajnantor plateau. Furthermore, conditions on the peaks surrounding the ALMA site are even better, particularly when thermal inversions trap much of the water vapor below the mountain summits. Under these circumstances, the transmission in the submillimeter (350 µm and 450 µm) windows is excellent and limited transmission exists even in windows up to 150 µm (Marrone et al. 2005).

Of the several peaks reasonably close to ALMA, Cerro Chajnantor (5600 m) has been selected as the candidate CCAT site. Other projects are also interested in this mountain, which lies within the recently expanded CONICYT science preserve, and the University of Tokyo has constructed a road to the summit area. For CCAT, the candidate location is a small shelf about 150 m northeast and 50 m below the summit ridge (Figure 2). This location is shielded from the prevailing westerly winds and is not visible from San Pedro. In 2006 May an autonomous instrument suite was deployed to assess observing conditions at the candidate CCAT site. The data are freely available (www.submm.org/site). Meteorological measurements show the candidate CCAT site is extremely dry and experiences typical wind speeds a bit lower than at the ALMA site, which is a pleasant surprise. Two tipping radiometers, one at the candidate CCAT site and one at the CBI on the ALMA plateau (5050 m) for reference, simultaneously measure the atmospheric transparency at 350 μm. These comparative measurements show the submillimeter transparency, which is largely determined by the water vapor column, is significantly better at the CCAT candidate site than at ALMA (Figure 3). Observing conditions at the CCAT site are excellent, almost as much better than ALMA as ALMA is better than Mauna Kea.

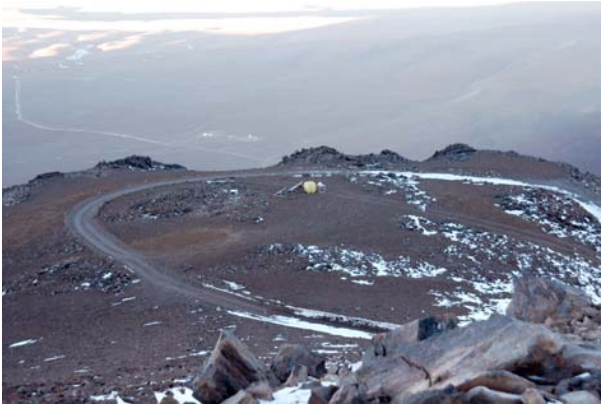

Fig. 2. View to the northeast over the CCAT site and meteorological equipment at 5612 m near the summit of Cerro Chajnantor. The existing ASTE and NANTEN2 telescopes are visible on the plateau 800 m below.

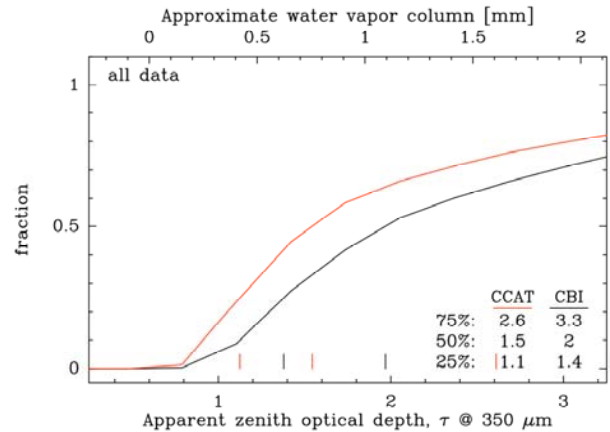

Fig. 3. Cumulative distributions of simultaneous measurements of the apparent submillimeter (350 μm) optical depth at the CCAT candidate site (5610 m) and at the CBI (5050 m) for 2006 May–2007 April. The CCAT site enjoys significantly better conditions than the CBI or ALMA, almost as much better as ALMA enjoys compared with Mauna Kea.

## 4. TELESCOPE DESIGN

For CCAT, the primary technical requirements are a large aperture (25 m), a wide field of view (20'), high efficiency at submillimeter (> 200 μm) wavelengths, accurate offset pointing (better than 0.3" rms), and agile scanning performance. High aperture efficiency demands high quality optics, namely a half wavefront error ≤ 10 μm rms for all the principal optical surfaces together. Relative to its size, CCAT requires an optical quality substantially better than any existing radio telescope. Fundamental limits on specific stiffness and thermal stability mean this level of performance is not likely possible with a passive telescope design, even with a carbon fiber structure. To meet this challenge, therefore, the CCAT design (Sebring et al. 2006a, 2006b) incorporates closed loop active control of the primary mirror, where the relative positions of the mirror panels are sensed and then controlled with actuators. Although a few radio telescopes (CSO, GBT, Noto) use open loop control to correct for gravitational distortions with look up tables and several optical telescopes (e. g., Keck) successfully use closed loop control of segmented primary mirrors, CCAT will be the first radio telescopes to implement a closed loop active primary mirror.

### 4.1 Optical design

For optimum performance over the full field of view, CCAT has a Ritchey Chrétien design (Cortés-Medellín & Herter 2006). To allow rapid exchanges, the instruments will be mounted at the Nasmyth foci behind the primary mirror and outside the elevation bearings. These foci are inside the outer diameter of the primary mirror to keep the secondary mirror size, the back focal distance, and the dome size small. Two bent Cassegrain foci are also available on the tipping structure for small science or telescope diagnosis instruments. The primary focal ratio, *f*/0.4, is typical for a radio telescope. Across the field of view, the image quality is excellent. Although the optimal focal surface is curved, short wavelength, wide field instruments can accommodate this curvature either with corrective coupling optics or with a segmented detector array.

### 4.2 Primary mirror

The primary mirror is comprised of 210 keystone shaped panels arranged in seven rings. The panels, with maximum dimensions of 1.6 m × 1.9 m, are kinematically mounted at just three points. To meet the telescope's overall wavefront error specification, the surface accuracy tolerance for the panels is ≤ 5 µm rms including manufacturing errors and thermal and gravitational deformations. Preliminary analysis indicates several sandwich technologies can meet the requirements, including carbon fiber skins with aluminum honeycomb cores (Martin et al. 2006), molded lightweight borosilicate glass (Strafford et al. 2006), and electroformed nickel skins with aluminum honeycomb cores (Valsecchi 2003). In all these technologies, the reflecting surfaces are replicated from precision mandrels and, in the cases of carbon fiber or glass, coated with aluminum. A bolted truss, most likely made of steel for cost reasons, supports the primary mirror panels and actuators. Because such a truss is neither stiff nor thermally stable enough to maintain the necessary accuracy, active control of the panel positions is required.

### 4.3 Active mirror control

Although the success of the Keck and other optical telescopes shows closed loop active control of a segmented primary mirror is feasible, applying the technique in the technical and financial context of a radio telescope poses some challenges. Edge sensors will be used to measure the relative displacements and dihedral angles of adjacent primary mirror panels. These measurements are very effective for controlling small scale, panel to panel scale mirror distortions. Unfortunately, however, they are comparatively insensitive to large scale deformations of the entire telescope, such as focus, astigmatism, etc., that will be introduced by varying gravitational loads and thermal gradients. Moreover, because the edge sensors and actuators are physically displaced but coupled through the panels, thermal and gravitational distortions of individual panels, even if radiometrically insignificant, can adversely affect the overall mirror alignment. Hence supplemental metrology systems, such as laser rangefinders or angle sensors, are under consideration. Preliminary models indicate the active control system will work, although many details remain to be settled. Shearing interferometry observations of planets (Serabyn 2006), which have been successfully used at the CSO for many years, will be used for initial alignment of the mirror panels and for calibration of the active control system.

### 4.4 Secondary and tertiary mirror

The 2.6 m diameter secondary mirror and 1.9 m × 2.7 m tertiary mirror will be made with the same technology as the primary mirror panels. Because they are unique optics, they may be figured individually rather than replicated from a mandrel. The secondary mirror will be mounted on a five axis hexapod and will probably have a nutator for rapid beam switching during spectroscopic observations. The tertiary mirror rotates to direct the beam to any of the four foci.

### 4.5 Mount

Designs for the optics and the telescope mount were developed together. The concept mount design (Finley et al. 2006) has hydrostatic azimuth bearings with journals mounted on a cylindrical pier and radial runout controlled by a central pintle bearing. Two diametrically opposed pairs of geared motors drive against a large helical gear mounted on the inside diameter of the azimuth journal. The much lighter elevation stage uses rolling element bearings and a pair of geared motors driving a helical gear sector. To allow the full 20 arcmin field of view to pass through to the instruments, the elevations bearings have large (> 1 m diameter) apertures. Tape encoders with multiple read heads are used on both axes. Preliminary servo control models indicate the telescope will meet the stringent requirements for pointing accuracy and scanning motions. To meet the challenge of accurate pointing, an auxillary optical or near IR telescope will be used to measure the mount characteristics. Then submillimeter observations will establish the radiometric pointing. Closed loop guiding in the near IR with the auxillary telescope is also under consideration.

### 4.6 Dome

To reduce the adverse influence of wind loads and solar heating on telescope quality and pointing and to provide shelter for science instruments and telescope staff, CCAT will be enclosed in a dome. For a 25 m telescope, the smallest dome that permits free movement of the telescope inside is about 42 m in diameter. Parametric assessment of several dome types (Loewen et al. 2006) concluded a calotte design offers significant mechanical advantages and will be cheaper to build and to operate. The calotte design achieves all sky pointing with two rotation stages, a conventional azimuth stage parallel to the ground and a second stage inclined about 38° to the vertical. This design is balanced about both rotation axes and features a fixed circular aperture, which provides good protection from the wind. The nearly continuous spherical shell lends itself to efficient space frame construction. An interior shutter rotates independently to meet the aperture near the horizon. Compared with other dome types, the calotte design is lighter, exhibits less and more uniformly distributed stress throughout the structure, and is balanced about both axes.

**4.7 Facilities**

On the mountain, the telescope facility (Terán et al. 2006) includes a control room, an office area, a computer and electronics room, and an instrument preparation lab. Because of the high altitude, these working spaces will have oxygen enrichment to enhance the safety, comfort, and productivity of observers and telescope staff. Portable supplemental oxygen will be used for exterior work during operations and during all phases of construction. Separate outbuildings house mechanical equipment such as chillers, compressors, etc., to minimize vibration and heat loading in the telescope enclosure. A lower altitude (2500 m) support facility near San Pedro will provide housing for telescope staff as well as a remote control room, offices, and instrument workshops.

## 5. INSTRUMENTATION

In recent years, large format bolometer arrays for submillimeter astronomy have developed rapidly. CCAT is specifically designed to take advantage of these developments and the primary science objectives emphasize wide field imaging and surveys. Hence the initial plan for facility instruments (Stacey et al. 2006) features two large format bolometer cameras, one for submillimeter wavelengths, 200–620 μm, and the other for near millimeter wavelengths, 740–2000 μm. Spectrometers, both direct detection and heterodyne, are also under consideration and existing, previous generation instruments developed for other facilities, will be brought to CCAT to enhance the scientific yield of the cameras. Although existing instruments cannot accomplish all the primary CCAT science objectives, they will provide important supplementary capabilities, especially in the early years of operation. Looking to the future, foreseeable instrument developments will extend the CCAT science return for many years.

**5.1 Direct illumination cameras**

At short submillimeter wavelengths the leading camera technology is directly illuminated TES silicon bolometers, exemplified by SCUBA2 (Holland et al. 2006b), now approaching initial deployment on the JCMT. When complete, SCUBA2 will simultaneously observe two bands, 450 μm and 850 μm, with 5120 pixels each. On CCAT, SCUBA2 would Nyquist sample a 2.7′ × 2.7′ field of view at 450 μm. By the time CCAT begins observations, SCUBA2 will have been in use for several years and will be a proven instrument, making it ideal for telescope commissioning. In addition, SCUBA2 would take advantage of the excellent observing conditions at CCAT to make very valuable science observations, in its own right, especially at 450 μm.

To follow SCUBA2, the concept for a short wavelength camera (SWCam) has 32,000 directly illuminated TES silicon bolometers spaced to Nyquist sample a 5′ × 5′ field of view (FoV) at 350 μm. The detector array is made of 25 edge butted 32 × 40 pixel subarrays of the type now produced for SCUBA2. Mesh filters, well matched to the atmospheric windows and mounted in a wheel immediately behind the Lyot stop, select the observing wavelength. Transmissive optics with diamond antireflection coated germanium lenses couple the array to the telescope focus. To ease the operations support requirements, the camera will use closed cycle cryogenics with pulse tube coolers followed by $^4$He and $^3$He or ADR stages.

**5.2 Antenna coupled cameras**

For wavelengths beyond the niobium superconducting energy gap (2Δ ≈ 725 GHz ≈ 410 μm), it is possible to construct antenna coupled pixels with lithographed superconducting microstrip. These phased array pixels allow precise beam formation and simultaneous multicolor observation. Recently a small camera demonstrating this technique has been deployed to the CSO (Schlaerth et al. 2007). In the long wavelength camera (LWCam) concept, each pixel has 16 slot dipole antennae with 16 taps on each slot. The signals are summed in a binary tree combiner and microstrip bandpass filters separate the frequency bands, providing simultaneous multicolor observations. For detectors, both TES bolometers and MKID resonators are under consideration. In total, there are 45,000 pixels with the central 10′ field Nyquist sampled at 850 μm and the entire 20′ × 20′ field of view covered at the longer wavelengths. To insure minimum optical loading, the camera is coupled to the telescope with reflective optics.

**5.3 Direct detection spectrometers**

Moderate spectral resolution, broadband, direct detection spectrometers would be valuable on CCAT for determining the redshift of distant galaxies, for example, or for line surveys in nearby galaxies. A current instrument is Zspec (Erle et al.

2007), which uses the WaFIRS architecture with a Roland grating in a parallel plate waveguide and an array of bolometers to achieve a resolving power of 200–400 over the entire 185–305 GHz atmospheric window. For CCAT, a similar design could cover the 350 µm and 450 µm windows simultaneously. Moreover, laboratory studies of flexible dielectric waveguides suggest multiobject instruments are possible, albeit with considerable development.

### 5.4 Heterodyne cameras

Novel packaging techniques and advances in digital signal processing now permit the construction of moderate scale heterodyne spectrometer arrays, such as the 64 pixel, 350 GHz SuperCam (Groppi et al. 2007). For CCAT, a preliminary concept is a 128 pixel, 650 GHz array with a 2–4 GHz bandwidth.

## 6. CONSORTIUM AND SCHEDULE

In 2005, Caltech, Cornell, and JPL carried out the CCAT feasibility and concept design study. Concurrently, the UK Astronomy Technology Centre proposed a similar concept (Holland et al. 2006a). At present these institutions, along with the University of Colorado and the Universities of British Columbia and of Waterloo, are establishing a CCAT consortium to pursue construction and operation of the telescope. Other institutions have also expressed interest in the project. The present development schedule aims for initial observations in 2012 and full operation in 2014, concurrent with the start of full ALMA operations.

## 7. SUMMARY

The Cornell Caltech Atacama Telescope will be a significant new facility for submillimeter astronomy, especially for sensitive wide field surveys. The high, dry site enjoys superb observing conditions. The telescope's combination of large aperture, high quality optics, and wide field will be unequalled. CCAT will provide a platform for instruments that take maximum advantage of rapidly developing detector technology. The concept telescope design builds on successful prior telescopes and the proposed technologies are largely within the state of the art. Several institutions are establishing a consortium to construct and operate the telescope.